\documentclass[aps,prd,reprint,groupedaddress,amsfonts]{revtex4-1}

\usepackage[utf8]{inputenc}
\usepackage[OT4]{fontenc}

\usepackage{bm}
\usepackage{graphicx}

\usepackage{physics}
\usepackage[hidelinks]{hyperref}

\begin{document}

\title{Spatial entanglement of nonvacuum Gaussian states}

\author{Filip Kiałka}
\email{fk322204@okwf.fuw.edu.pl}
\altaffiliation{Present address: University of Duisburg-Essen, Duisburg, Germany.}
\author{Mehdi Ahmadi}
\email{mehdi.ahmadi@ucalgary.ca}
\altaffiliation{Present address: University of Calgary, Calgary, AB, Canada.}
\author{Andrzej Dragan}
\email{dragan@fuw.edu.pl}

\affiliation{Institute of Theoretical Physics, Faculty of Physics, University of Warsaw, Pasteura 5, 02-093 Warsaw, Poland}

\date{\today}

\begin{abstract}
The vacuum state of a relativistic quantum field contains entanglement between regions separated by spacelike intervals. Such spatial entanglement can be revealed using an operational method introduced in~\cite{Rodriguez-VazquezEtAl2014,BrownEtAl2014}. In this approach, a cavity is instantaneously divided into halves by an introduction of an extra perfect mirror. Causal separation of the two regions of the cavity reveals nonlocal spatial correlations present in the field, which can be quantified by measuring particles generated in the process. We use this method to study spatial entanglement properties of nonvacuum Gaussian field states. In particular, we show how to enhance the amount of harvested spatial entanglement by an appropriate choice of the initial state of the field in the cavity. We find a counterintuitive influence of the initial entanglement between cavity modes on the spatial entanglement which is revealed by dividing the cavity in half.
\end{abstract}

\maketitle

\section{\label{sec:introduction} Introduction} 

The vacuum fluctuations of a relativistic quantum field in spacelike separated regions are quantum correlated. This is referred to as vacuum entanglement. It has been predicted in the context of accelerating frames of reference~\cite{Fulling1973,Unruh1976}, as well as in algebraic quantum field theory~\cite{SummersWerner1985,SummersWerner1987a,SummersWerner1987b,SummersWerner1987}. In~\cite{Reznik2000,Reznik2003}, it has been shown how entanglement can be extracted from the vacuum state of a quantum field using a pair of initially unentangled two-level atoms, known as Unruh-DeWitt detectors, which interact only locally with the field. A more detailed analysis of this scenario was later performed in~\cite{LinHu2010}. Moreover, a feasible method for detecting vacuum entanglement was proposed in a chain of trapped ions~\cite{RetzkerEtAl2005}. Such entanglement is believed to persist between arbitrarily far away regions of bosonic and fermionic fields~\cite{Redhead1995,VerchWerner2005,ReznikEtAl2005,SilmanReznik2007}, and it manifests in the Unruh effect for accelerating observers~\cite{Reznik2003,LinHu2010,DraganEtAl2013,Dragan2013PRA,DoukasEtAl2013}. Recently, however, a thorough analysis of vacuum entanglement was performed for massive fields~\cite{AhmadiEtAl2016}, where sudden death of vacuum entanglement as a function of the distance between the modes was observed. It has also been shown that vacuum entanglement can be used to violate Bell's inequalities~\cite{SummersWerner1985,SummersWerner1987a,SummersWerner1987b,SummersWerner1987,ReznikEtAl2005}, can be multipartite~\cite{SilmanReznik2005,LorekEtAl2014}, and that for massless fields, it persists between timelike separated regions~\cite{OlsonRalph2011}. It can also be resonantly enhanced with the use of a moving cavity \cite{BruschiEtAl2013}. Furthermore, the phenomenon is expected to be sensitive to spacetime curvature~\cite{ClicheKempf2011,SteegMenicucci2009} and to have a relevant effect on the possibility of building ideal relativistic clocks \cite{LorekEtAl2015,Lindkvist2014PRA}.

In this paper, we generalize one of the recent operational methods for investigating vacuum entanglement~\cite{Rodriguez-VazquezEtAl2014,BrownEtAl2014} to nonvacuum Gaussian states. In algebraic quantum field theory, it has been stressed that essentially all field states possess the aforementioned spatial entanglement properties~\cite{SummersWerner1987,CliftonHalvorson2001}. However, to the best of our knowledge, states other than the vacuum state and the thermal state~\cite{NgBurnett2007,Brown2013} have not yet been explored in the operational approach. In particular, it is not known whether nonclassical states, such as squeezed states, offer any experimental advantage. Also, it is not clear how the entanglement between spatial areas and the entanglement in the global basis of plane waves or standing waves are related to each other. 

In this paper, we address the above questions. Specifically, we consider coherent states, single-mode squeezed thermal states, and two-mode squeezed vacuum states. We show that spatial entanglement is indeed not unique to vacuum. Moreover, we provide examples of states for which the effect is stronger and hence easier to detect. We also observe that the amount of entanglement in the usual standing wave basis can contribute to the amount of spatial entanglement.

In our operational approach, an optical cavity is split into two smaller cavities via rapid introduction of a mirror in between the two mirrors of the cavity. We show that particle production due to such a change of boundary conditions can be made stronger by using a nonvacuum initial state of the field inside the cavity. Particle creation due to rapidly varying boundary conditions is of independent interest, and our work adds to the body of theoretical predictions concerning this phenomenon~\cite{LozovikEtAl1995,ObadiaParentani2001,FedotovEtAl2005,Rodriguez-VazquezEtAl2014,BrownLouko2015}.

This paper is organized in the following way. In Sec.~\ref{sec:methods}, we briefly review the operational model for studying vacuum entanglement. In Sec.~\ref{sub:coherent_states}, we focus on the particle creation effect for initial coherent states. Then, we study spatial entanglement and particle production for a single-mode squeezed thermal state in Sec.~\ref{sub:one_mode_squeezed_thermal_states}. We devote Sec.~\ref{sub:two_mode_squeezed_vacuum} to investigating whether the final amount of spatial entanglement between the localized modes is affected by the presence of entanglement between the initial modes of the cavity. Finally, in Sec.~\ref{sec:conclusions}, we present our conclusions, open questions and possible future lines of research. 

\begin{figure}[t]
	\includegraphics[width=0.47\textwidth]{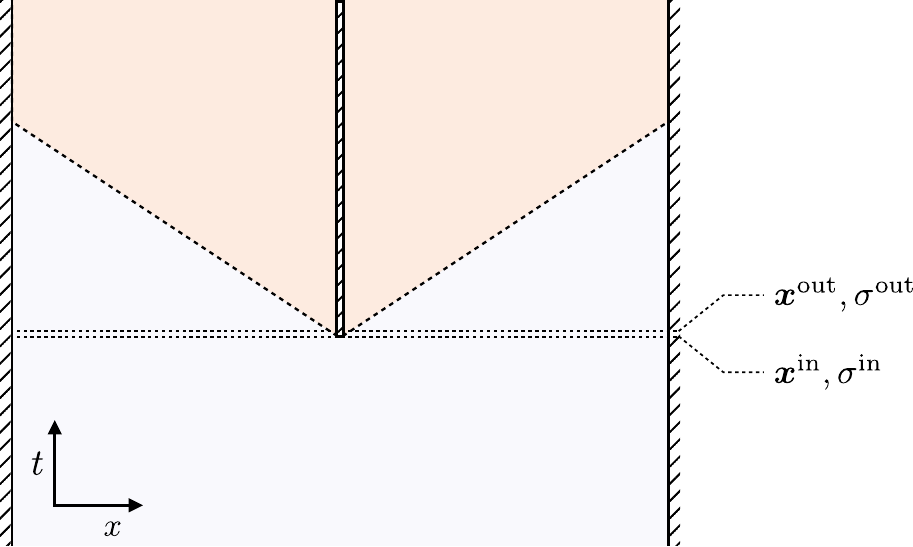}
	\caption{\label{fig:setup}
		We investigate how the state of a quantum field in a cavity changes when the cavity is instantaneously divided in half with an additional mirror. After such a rapid division quantum correlated particles are produced~\cite{Rodriguez-VazquezEtAl2014,BrownEtAl2014}. The input and output states are Gaussian, characterized by the vectors of first moments $\bm{x}^\text{in}$, $\bm{x}^\text{out}$ and the covariance matrices $\sigma^\text{in}$, $\sigma^\text{out}$.}
\end{figure}

\section{\label{sec:methods} Model} 
In this section, we briefly review the operational approach of~\cite{Rodriguez-VazquezEtAl2014,BrownEtAl2014}. Consider a one-dimensional Dirichlet cavity, in the middle of which an additional mirror is instantaneously introduced (see Fig.~\ref{fig:setup}). The cavity contains a massless scalar field in an initial Gaussian state. First, we calculate the state of the field after the mirror was introduced. This way we are able to investigate the mean number of particles produced due to introduction of the mirror. Moreover, we analyze the amount of entanglement between different modes of the two smaller cavities as a function of the initial state of the field.

To show the correspondence between entanglement of particles created in this model and vacuum entanglement, we recall an extended scheme due to Brown \emph{et al.}~\cite{BrownEtAl2014}. In this scheme two mirrors are simultaneously introduced a finite distance apart, and it is found that particles created by one mirror are entangled with those created by the other. Since the particle creation events are spacelike separated, one argues that the obtained amount of entanglement must have been present in the initial state of the field. In this paper, we use this scheme, but in a limit where distance between the two mirrors approaches zero. Nevertheless, we can retain the interpretation of extracting entanglement from the initial state rather than creating~it.

\subsection{\label{sub:classical_solutions_and_the_bogoliubov_transformation} Classical solutions and the Bogoliubov transformation} 

	Consider a one-dimensional cavity with a massless scalar field, described by the Klein-Gordon equation:
		\begin{equation} \label{eq:K-G}
			\qty(\partial^2_t - \partial^2_x) \phi\qty(t,x) = 0,
		\end{equation}
	and Dirichlet boundary conditions $\phi\qty(t,0) = \phi\qty(t,R) = 0$ for all $t$. At $t=0$, an additional mirror is introduced at $x=r$, which corresponds to the condition $\phi\qty(t,r) = 0$ for $t > 0$. In Sec.~\ref{sec:results}, we will assume $r = R/2$, in this section; however, we review the transformation procedure for arbitrary $r \in \qty[0,R]$ as given in~\cite{Rodriguez-VazquezEtAl2014,BrownEtAl2014}.

	We introduce two families of solutions of the Klein-Gordon equation~\eqref{eq:K-G}. The first family, $\qty(U_l)_{l \in \mathbb{N}}$, physically corresponds to the standing waves before the mirror is inserted. The second family of solutions are $\qty(u_n)_{n \in \mathbb{N}}$, $\qty(\bar{u}_{n})_{n \in \mathbb{N}}$. These are the standing waves contained within the left cavity and within the right cavity after the division, respectively. They read as
	\begin{subequations} \label{eq:U_l}
	\begin{eqnarray}
		U_l\qty(t,x) &\overset{t < 0}{=}& \frac{1}{\sqrt{R \Omega_l}} \sin(\frac{\pi l x}{R}) \mathrm{e}^{-\mathrm{i} \Omega_l t},\\
		\Omega_l &=& \frac{\pi l}{R},
	\end{eqnarray}
	\end{subequations}
	and
	\begin{subequations} \label{eq:u_n}
	\begin{eqnarray}
		u_n\qty(t,x) &\overset{t > 0}{=}& \frac{\theta\qty(r - x)}{\sqrt{r \omega_n}} \sin(\frac{\pi n x}{r}) \mathrm{e}^{-\mathrm{i} \omega_n t},\\
		\omega_n &=& \frac{\pi n }{r},\\
		\bar{u}_n\qty(t,x) &\overset{t > 0}{=}& \frac{\theta\qty(x - r)}{\sqrt{\qty(R-r) \bar{\omega}_n}} \sin(\frac{\pi n \qty(x - r)}{R - r}) \mathrm{e}^{-\mathrm{i} \bar{\omega}_n t},\\
		\bar{\omega}_n &=& \frac{\pi n }{R - r},
	\end{eqnarray}
	\end{subequations}
	where $\theta\qty(x)$ is the Heaviside step function. Having introduced the two families of solutions, we can relate them by a Bogoliubov transformation:
	\begin{subequations} \label{eq:bogotrans}
	\begin{eqnarray}
		u_n\qty(x,t) &=& \sum_{n=1}^\infty \qty( \alpha_{nl} U_l\qty(x,t) + \beta_{nl} U^*_l\qty(x,t) ),\\
		\bar{u}_n\qty(x,t) &=& \sum_{n=1}^\infty \qty( \bar{\alpha}_{nl} U_l\qty(x,t) + \bar{\beta}_{nl} U^*_l\qty(x,t) ).
	\end{eqnarray}
	\end{subequations}
	The coefficients $\alpha_{nl},\,\beta_{nl},\,\bar{\alpha}_{nl},\,\bar{\beta}_{nl}$ are given by Klein-Gordon scalar products~\cite{BirrellDavies1986} of solutions $U_l$ with $u_n$, $\bar{u}_n$, i.e., the overlaps of the mode functions corresponding to input states with the mode functions associated with the output states. To calculate these overlaps, we need to extend $U_l$ past the $t = 0$ point so that there is a Cauchy surface on which both families of solutions are defined. This would normally require solving equation~\eqref{eq:K-G} with a time-dependent boundary condition at the point where the mirror is introduced. However, for infinitely fast introduction of the mirror the dynamics can be neglected. Therefore, we simply extend $U_l$ using free evolution of the field and then we calculate the scalar products, which leads to these Bogoliubov coefficients:
	\begin{subequations} \label{eq:bogocoeffs}
	\begin{eqnarray}
		\alpha_{nl} &=& \left( U_l \middle | u_n \right) = \qty(\Omega_l + \omega_n) \mathcal{V}_{nl},\\
		\beta_{nl} &=& - \left( U^*_l \middle | u_n \right) = \qty(\Omega_l - \omega_n) \mathcal{V}_{nl},\\
		\bar{\alpha}_{nl} &=& \left( U_l \middle | \bar{u}_n \right) = \qty(\Omega_l + \bar{\omega}_n) \bar{\mathcal{V}}_{nl},\\
		\bar{\beta}_{nl} &=& - \left( U^*_l \middle | \bar{u}_n \right) = \qty(\Omega_l - \bar{\omega}_n) \bar{\mathcal{V}}_{nl},
	\end{eqnarray}
	\end{subequations}
	where
	\begin{subequations} \label{eq:nus}
	\begin{eqnarray}
		\mathcal{V}_{nl} &=& \begin{cases}
			\frac{ (-1)^n n \pi }{r \sqrt{R r \Omega_l \omega_n} \qty(\Omega_l^2 - \omega_n^2)} \sin(\frac{l \pi r}{R}) & \Omega_l \ne \omega_n, \\
			\frac{r}{2 \sqrt{R r \Omega_l \omega_n}} & \Omega_l = \omega_n,
		\end{cases} \\
		\bar{\mathcal{V}}_{nl} &=& \begin{cases}
			\frac{ - n \pi }{\bar{r} \sqrt{R \bar{r} \Omega_l \bar{\omega}_n} \qty(\Omega_l^2 - \bar{\omega}_n^2)} \sin(\frac{l \pi r}{R}) & \Omega_l \ne \bar{\omega}_n, \\
			\frac{(-1)^{n+l} \bar{r}}{2 \sqrt{R \bar{r} \Omega_l \bar{\omega}_n}} & \Omega_l = \bar{\omega}_n.
		\end{cases}
	\end{eqnarray}
	\end{subequations}
	
	At this point it is worth mentioning that a more careful calculation of these Bogoliubov coefficients has been performed in~\cite{BrownLouko2015}, which involves solving Eq.~\eqref{eq:K-G} with a time-dependent boundary condition. There, $\phi\qty(t,r) = 0$ is enforced with a boundary condition (for spatially even solutions),
	\begin{equation} \label{eq:jorma_bc}
		\lim_{x \rightarrow r^\pm} \frac{\partial_x \phi\qty(t,x)}{\phi\qty(t,x)} = \pm B\qty(t),
	\end{equation} 
	where $B\qty(t)$ smoothly changes from $0$ to infinity as the mirror is introduced. The obtained Bogoliubov coefficients for an evolution satisfying condition~\eqref{eq:jorma_bc} can be seen to reduce to the coefficients \eqref{eq:bogocoeffs} in the instantaneous mirror introduction limit. Thus, our approximation that neglects the dynamics is well founded.

	\subsection{\label{sub:quantum_model} Quantum model} 

	Let us proceed by briefly reviewing the quantized version of the model introduced in the previous section. In the canonical quantization procedure with each solution $\qty(U_l)_{l=1,\dots,2 \Lambda}$ (where $2 \Lambda$ is a UV cutoff) we associate a pair of time-independent hermitian quadrature operators $\hat{Q}_l,\,\hat{P}_l$. The quadrature operators obey canonical bosonic commutation relations:
	\begin{equation}
		\comm{\hat{Q}_i}{\hat{P}_{j}} = \mathrm{i} \delta_{ij}.
	\end{equation}
	Similarly, we associate quadrature operators $\hat{q}_n,\,\hat{p}_n$ and $\hat{\bar{q}}_m,\,\hat{\bar{p}}_m$ with the solutions $\qty(u_n)_{n=1,\dots,\Lambda}$ and $\qty(\bar{u}_m)_{m=1,\dots,\Lambda}$. We conveniently arrange these operators into two vectors:
	\begin{eqnarray}
		\hat{\bm{x}}^\text{in} &=& \qty(\hat{Q}_1, \hat{P}_1, \ldots, \hat{Q}_{2\Lambda}, \hat{P}_{2\Lambda}),\\
		\hat{\bm{x}}^\text{out} &=& \qty(\hat{q}_1, \hat{p}_1, \ldots, \hat{q}_{\Lambda}, \hat{p}_{\Lambda}, \hat{\bar{q}}_{1}, \hat{\bar{p}}_{1}, \ldots, \hat{\bar{q}}_{\Lambda}, \hat{\bar{p}}_{\Lambda}).
	\end{eqnarray}
	Then, the Bogoliubov transformation~\eqref{eq:bogotrans} can be written as
	\begin{equation} \label{eq:bogotrans_quadr}
		\hat{\bm{x}}^\text{out} = S \hat{\bm{x}}^\text{in},
	\end{equation}
	where $S$ is a square matrix:
	\begin{equation} \label{eq:S_block_form}
		S = \mqty( S_{1,1} & \cdots & S_{1,\,2\Lambda} \\
		\vdots & \ddots & \vdots \\
		S_{\Lambda,\,1} & \cdots & S_{\Lambda,\,2\Lambda} \\
		\bar{S}_{1,1} & \cdots & \bar{S}_{1,\,2\Lambda} \\
		\vdots & \ddots & \vdots \\
		\bar{S}_{\Lambda,\,1} & \cdots & \bar{S}_{\Lambda,\,2\Lambda} ),
	\end{equation}
	where the elements listed above are $2 \times 2$ matrices given by	\begin{subequations} \label{eq:S_blocks}
	\begin{eqnarray}
		S_{nl} &=& 2 \mathcal{V}_{nl} \mqty(\dmat[0]{\omega_n, \Omega_l}), \\
		\bar{S}_{nl} &=& 2 \bar{\mathcal{V}}_{nl} \mqty(\dmat[0]{\bar{\omega}_n, \Omega_l}).
	\end{eqnarray}
	\end{subequations}

	In this paper, we limit ourselves to Gaussian initial states~\cite{[{For a modern overview of Gaussian states and their manipulation see }]WeedbrookEtAl2011}. Since the Bogoliubov transformation~\eqref{eq:bogotrans_quadr} is linear in quadrature operators, the transformed state will also be Gaussian. Therefore, the first and second statistical moments are sufficient to characterize both the initial and final states, i.e., the state of the field before and after the introduction of the mirror. The first moments are given by a vector of expectation values,
	\begin{equation} \label{eq:def_fmvector}
		\bm{x} = \expval{\hat{\bm{x}}},
	\end{equation}
	while the second moments are given by a covariance matrix, which consists of $2 \times 2$ blocks defined in the following way:
	\begin{equation} \label{eq:def_covm}
		\sigma_{ij} = \mqty(\expval{\acomm{\Delta \hat{q}_i}{\Delta \hat{q}_j}} & \expval{\acomm{\Delta \hat{q}_i}{\Delta \hat{p}_j}} \\
		\expval{\acomm{\Delta \hat{p}_i}{\Delta \hat{q}_j}} & \expval{\acomm{\Delta \hat{p}_i}{\Delta \hat{p}_j}}),
	\end{equation}
	where $\Delta \hat{A} = \hat{A} - \big< \hat{A} \big>$ and $\acomm{}{}$ is the anticommutator. The Bogoliubov transformation~\eqref{eq:bogotrans_quadr} implies the following transformation laws for the vector of first moments and the covariance matrix hold:
	\begin{subequations}
	\begin{eqnarray}
		\bm{x}^\text{out} &=& S \bm{x}^\text{in}, \label{eq:trafo_fmvector} \\
		\sigma^\text{out} &=& S \sigma^\text{in} S^\mathrm{T}. \label{eq:trafo_covm}
	\end{eqnarray}
	\end{subequations}

	Finally, we are interested in two properties of the state after transformation, namely entanglement and the average number of particles. To calculate these, we divide the $4 \Lambda \times 4 \Lambda$ covariance matrix $\sigma^\text{out}$ into $2\Lambda \times 2\Lambda$ blocks:
	\begin{equation} \label{eq:sigma_out_block_form}
		\sigma^\text{out} = \mqty( \sigma & \gamma \\ \gamma^\mathrm{T} & \bar{\sigma} ).
	\end{equation}
	Then, we divide the blocks $\sigma$, $\gamma$, and~$\bar{\sigma}$ further into $2 \times 2$ blocks $\sigma_{ij}$, $\gamma_{ij}$, and~$\bar{\sigma}_{ij}$, describing the states of individual modes and correlations between them. In this block notation, we can easily give the expectation value of the particle number operator, for example for mode $u_n$, as
	\begin{equation} \label{eq:particle_number_formula}
		\expval{\hat{n}_n} = \frac{1}{4} \qty( \Tr \sigma_{nn} - 2 ) + \frac{1}{2} \qty( \expval{\hat{q}_n}^2 + \expval{\hat{p}_n}^2 ).
	\end{equation}
	While for mode $\bar{u}_n$ the formula contains the barred counterparts $\bar{\sigma}_{nn}$, $\hat{\bar{q}}_n$, and $\hat{\bar{p}}_n$. 
	
	To calculate the entanglement between a pair of modes $u_n$ and $\bar{u}_m$ we need to find the corresponding covariance matrix, which we will denote by $\sigma^\text{out}\vert_{nm}$. It is obtained by deleting all the entries of the matrix $\sigma^{\text{out}}$ except for the $16$ entries which lie on the intersection of the four rows and four columns corresponding to modes $u_n$ and $\bar{u}_m$:
	\begin{equation} \label{eq:alpha}
		\sigma^\text{out}\vert_{nm} = \mqty(\sigma_{nn} & \gamma_{nm} \\ \gamma_{nm}^\mathrm{T} & \bar{\sigma}_{mm}).
	\end{equation}
	We use logarithmic negativity~\cite{AdessoIlluminati2007} as an operational measure of entanglement between the modes, as it is an entanglement monotone and provides an upper bound to distillable entanglement. The logarithmic negativity of the reduced state of modes $u_n$ and $\bar{u}_m$ is easy to compute and is given by
	\begin{equation} \label{eq:negativity}
		E_\mathcal{N}\qty(n,m) = \max \qty{0,\,-\log \sqrt{ \frac{\tilde{\Delta} - \sqrt{\tilde{\Delta}^2 - 4 \det \sigma^\text{out}\vert_{nm}}}{2} }},
	\end{equation}
	where
	\begin{equation}
		\tilde{\Delta} = \det \sigma_{nn} + \det \bar{\sigma}_{mm} - 2\det \gamma_{nm}.
	\end{equation}

	\subsection{\label{sub:transformation_of_non_vacuum_states} Transformation of the covariance matrices of nonvacuum states} 
	
	The covariance matrix of the vacuum is unity, hence the result of the transformation~\eqref{eq:trafo_covm} for the vacuum as the initial state is simply~$\sigma^\text{out} = S S^\mathrm{T}$. Let us now assume the input state is the vacuum in all modes except $U_k$, in which we have a single-mode Gaussian state described by a covariance matrix~$\sigma^\text{in}_{kk}$. Under such circumstances, the $2 \times 2$ blocks of $\sigma$, $\gamma$, and~$\bar{\sigma}$ [see Eq.~\eqref{eq:alpha}], are equal to
	\begin{subequations} \label{eq:sigma_out_om_blocks}
	\begin{eqnarray}
		\sigma_{ij} &=& S_{ik} \sigma^\text{in}_{kk} S_{jk} + \sum^{2\Lambda}_{l \ne k} S_{il} S_{jl}, \\
		\gamma_{ij} &=& S_{ik} \sigma^\text{in}_{kk} \bar{S}_{jk} + \sum^{2\Lambda}_{l \ne k} S_{il} \bar{S}_{jl}, \\
		\bar{\sigma}_{ij} &=& \bar{S}_{ik} \sigma^\text{in}_{kk} \bar{S}_{jk} + \sum^{2\Lambda}_{l \ne k} \bar{S}_{il} \bar{S}_{jl}.
	\end{eqnarray}
	\end{subequations}
	The above result uses the fact that $S_{nl}$ are symmetric and reduces to~$S S^\mathrm{T}$ when~$\sigma^\text{in}_{kk} = \openone$.

	The second type of input state that is of interest in this paper is the one that has all modes in the vacuum state except for the two modes $U_k$, $U_{k'}$ ($k < k'$). The reduced state of these two modes is a two-mode Gaussian state described by a $4 \times 4$ covariance matrix:
	\begin{equation} \label{eq:init_two_mode_covm_blocks}
		\sigma^\text{in}\vert_{kk'} = \mqty(\sigma^\text{in}_{kk} & \sigma^\text{in}_{kk'} \\ \sigma^\text{in}_{k'k} & \sigma^\text{in}_{k'k'}).
	\end{equation}
	In this case, using Eqs.~\eqref{eq:S_blocks} and~\eqref{eq:trafo_covm}, the blocks of $\sigma$, $\gamma$, and~$\bar{\sigma}$, as defined in Eq.~\eqref{eq:alpha}, can be written as
	\begin{subequations} \label{eq:sigma_out_tm_blocks} 
	\begin{eqnarray}
		\sigma_{ij} &=& S_{ik} \, \sigma^\text{in}_{kk} \, S_{jk} +
		S_{ik'} \, \sigma^\text{in}_{k'k'} \, S_{jk'} +
		S_{ik} \, \sigma^\text{in}_{kk'} \, S_{jk'} + \dots \nonumber\\
		&&+ S_{ik'} \, \sigma^\text{in}_{k'k} \, S_{jk} +
		\sum^{2\Lambda}_{l \ne k,k'} S_{il} S_{jl},
	\\
		\gamma_{ij} &=& S_{ik} \, \sigma^\text{in}_{kk} \, \bar{S}_{jk} +
		S_{ik'} \, \sigma^\text{in}_{k'k'} \, \bar{S}_{jk'} +
		S_{ik} \, \sigma^\text{in}_{kk'} \, \bar{S}_{jk'} + \dots \nonumber\\
		&&+ S_{ik'} \, \sigma^\text{in}_{k'k} \, \bar{S}_{jk} +
		\sum^{2\Lambda}_{l \ne k,k'} S_{il} \bar{S}_{jl},
	\\
		\bar{\sigma}_{ij} &=& \bar{S}_{ik} \, \sigma^\text{in}_{kk} \, \bar{S}_{jk} +
		\bar{S}_{ik'} \, \sigma^\text{in}_{k'k'} \, \bar{S}_{jk'} +
		\bar{S}_{ik} \, \sigma^\text{in}_{kk'} \, \bar{S}_{jk'} + \dots \nonumber\\
		&&+ \bar{S}_{ik'} \, \sigma^\text{in}_{k'k} \, \bar{S}_{jk} +
		\sum^{2\Lambda}_{l \ne k,k'} \bar{S}_{il} \bar{S}_{jl}.
	\end{eqnarray}
	\end{subequations}

\section{\label{sec:results} Spatial entanglement and particle production} 

In this section, we first study particle production and entanglement extraction for coherent states and general single-mode Gaussian states as initial states of the large cavity before division. Then, we compare the spatial entanglement of a two-mode squeezed state and a state consisting only of its thermal marginals. This way, we are able to demonstrate that the result of introducing the mirror is sensitive to correlations present in the initial state of the cavity. Throughout this section, we assume that the additional mirror is introduced precisely in the middle of the cavity, i.e., $r = R/2$.

	\subsection{\label{sub:coherent_states} Coherent states} 

	For the first input state, we assume that all the modes are in the vacuum state, except for the mode $U_k$ which is prepared in a coherent state with the amplitude $\rho > 0$ and phase $\varphi$. Using Eqs. \eqref{eq:def_fmvector} and \eqref{eq:def_covm}, the vector of first moments and the covariance matrix of this state can be written, respectively, as
	\begin{subequations} \label{eq:init_coherent}
	\begin{eqnarray}
		x^\text{in}_n &=& \begin{cases}
			\rho \cos \varphi & n = k,\\
			\rho \sin \varphi & n = k +1,\\
			0 & \text{otherwise},
		\end{cases} \label{eq:init_coherent_fmoments}\\
		\sigma^\text{in} &=& \openone.
	\end{eqnarray}
	\end{subequations}

	At this point we note that changing the first moments of the initial state does not change the covariance matrix, and hence the entanglement, of the state after the introduction of the mirror [see Eqs.~\eqref{eq:trafo_covm} and~\eqref{eq:negativity}]. Since coherent states differ from the vacuum only in their first moments, their spatial entanglement is the same as that of the vacuum, which was already discussed elsewhere~\cite{Rodriguez-VazquezEtAl2014,BrownEtAl2014}. Therefore, in this subsection, we focus only on the average number of particles produced due to division of the cavity.
	
	\begin{figure}[t]
		\includegraphics[width=0.47\textwidth]{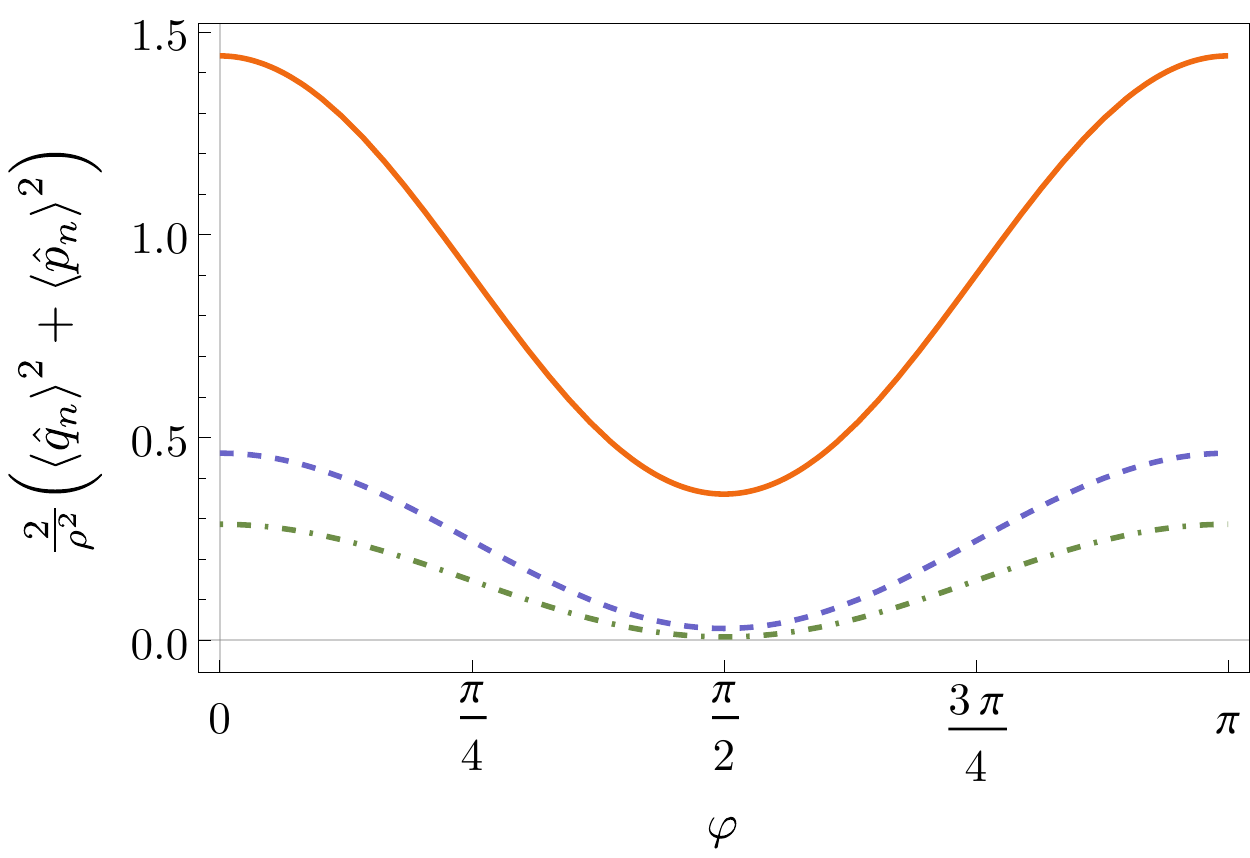}
		\caption{\label{fig:coherent_particles}
			Expected number of particles after a rapid division of the cavity containing a coherent state~\eqref{eq:init_coherent} is equal to a constant plus the term~\eqref{eq:res_coherent_particles} which is proportional to the initial number of particles. Here, the proportionality constant is shown as a function of phase for a coherent state initially in the first mode of the cavity. Solid, dashed and dot-dashed lines stand for the sum of the average numbers of particles in modes $u_n$ and $\bar{u}_n$, where $n=1,\,2,\,3$, respectively. Notice that the number of particles can either increase or decrease depending on phase~$\varphi$ of the initial coherent state.}
	\end{figure}

	The average number of particles after the introduction of the mirror is a sum of contributions from the first and from the second moments, as given in Eq.~\eqref{eq:particle_number_formula}. For a coherent state, the term corresponding to second moments is equal to the number of particles produced by inserting the mirror when the cavity is in the vacuum state. This is, however, a negligible number~\cite{Rodriguez-VazquezEtAl2014,BrownEtAl2014}. Since we are interested in the possible advantage of a coherent state over the vacuum state in the context of particle production, we need to discuss the contribution of the nonzero first moments. Using Eqs.~\eqref{eq:trafo_fmvector} and \eqref{eq:init_coherent_fmoments}, it is straightforward to calculate the expectation values appearing in Eq.~\eqref{eq:particle_number_formula} for the state of the mode $u_n$:
	\begin{equation} \label{eq:res_coherent_particles}
		\expval{\hat{q}_n}^2 + \expval{\hat{p}_n}^2 = 4 \rho^2 \mathcal{V}_{nk}^2 \qty(\omega_n^2 \cos^2 \varphi + \Omega_k^2 \sin^2 \varphi).
	\end{equation}
		
	Since we introduce the mirror precisely in the middle of the cavity, an identical result holds for the mode $\bar{u}_n$ on the right side of the mirror. As expected from the linearity of the transformation~\eqref{eq:bogotrans_quadr}, we observe that the final and initial average number of particles are proportional to each other. The proportionality constant is plotted in Fig.~\ref{fig:coherent_particles} for $k = 1$.

	The dependence of particle production on the phase $\varphi$ can be understood in terms of the expectation value of the field at the event where the mirror is inserted. We can see from Eqs.~\eqref{eq:init_coherent_fmoments} and~\eqref{eq:U_l} that $\varphi$ equal to~$0$ or~$\pi$ corresponds to states with the maximum value of $\big<\hat{\phi}\qty(0,r)\big>$, for which the amount of particles produced and the increase of energy are also maximal. The only choice of phase for which the energy of the state remains unchanged is $\varphi = \pi/2$. This corresponds to an initial state for which $\big<\hat{\phi}\qty(0,r)\big> = 0$. In this case, since the modes $u_n$ and $\bar{u}_n$ have higher frequencies than $U_l$ modes, energy conservation implies a decrease in the number of particles. This explains the absorption, rather than production, of particles we observe around $\varphi = \pi/2$.
	
	\begin{figure}[t]
		\centering
		\includegraphics[width=0.47\textwidth]{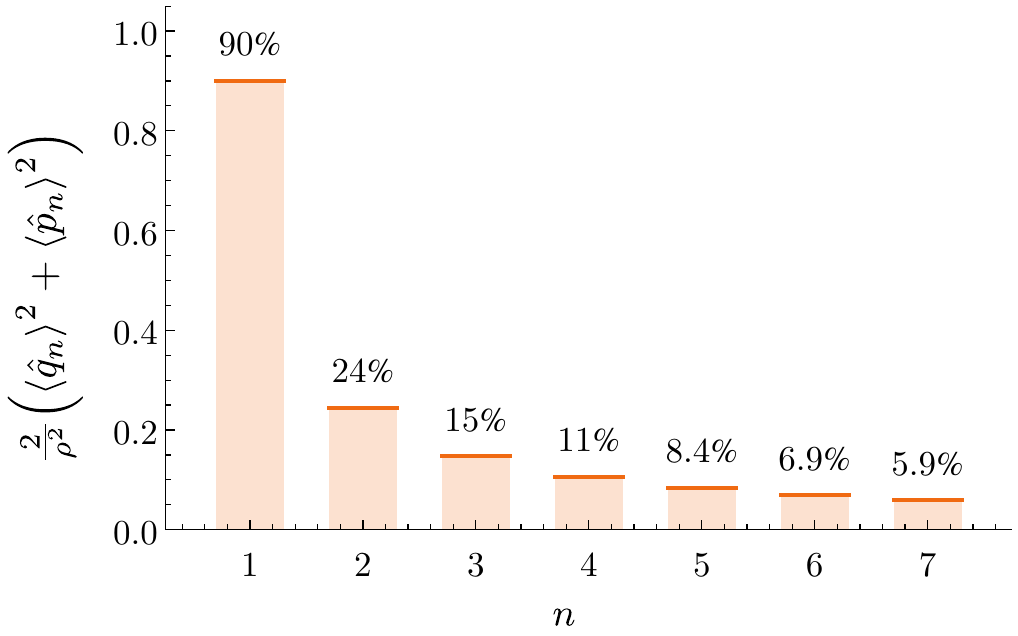}
		\caption{\label{fig:coh_average_particles}
			Results from Fig.~\ref{fig:coherent_particles} averaged over phase $\varphi$ of the initial coherent state~\eqref{eq:init_coherent}. The expected number of particles after inserting the mirror is expressed here as a percentage of the number of particles initially in the cavity. We see that the random phase of the initial coherent state implies at least two lowest modes need to be considered to witness the overall increase in the total number of particles.}
	\end{figure}
	
	From a quantum-optical perspective the dependence on phase, which is visible in Fig.~\ref{fig:coherent_particles}, resembles a degenerate parametric amplifier~\cite{WallsMilburn2008}. For the amplifier, however, the strength of particle production depends on the phase difference between the pumping and the amplified beam, whereas in our case no system providing the reference phase seems to be present. It turns out, however, that the phase reference is in fact provided in our setting by the choice of the boundary conditions on the mirror. Dirichlet boundary conditions, requiring that the field modes vanish at the mirror, do violate the phase symmetry of the field. Alternatively, forcing $\partial_t \phi$ to be zero at the mirror location would amplify coherent states with $\varphi = \pi/2$ rather than those with $\varphi = 0$ as it is in our case.
	
	Finally, if the phase of the initial coherent state is undetermined, then we need to consider a phase-averaged coherent state, which is non-Gaussian. Luckily, the particle number is a linear function of the density matrix, and thus the phase averaging of the initial state is equivalent to the phase averaging of the result~\eqref{eq:res_coherent_particles}. Figure~\ref{fig:coh_average_particles} shows the mean numbers of particles after introduction of a mirror into a cavity containing a phase-averaged coherent state. We observe that the average number of particles decays with frequency, and that the detection of particles in the few lowest modes is enough to witness significant particle production. We point out that the average particle number in modes $u_n$ or $\bar{u}_n$ decays slower than $1/n$, which causes the total number of particles to diverge. We postpone, however, the discussion of this divergence until the end of this section.

	\begin{figure}[t]
		\includegraphics[width=0.47\textwidth]{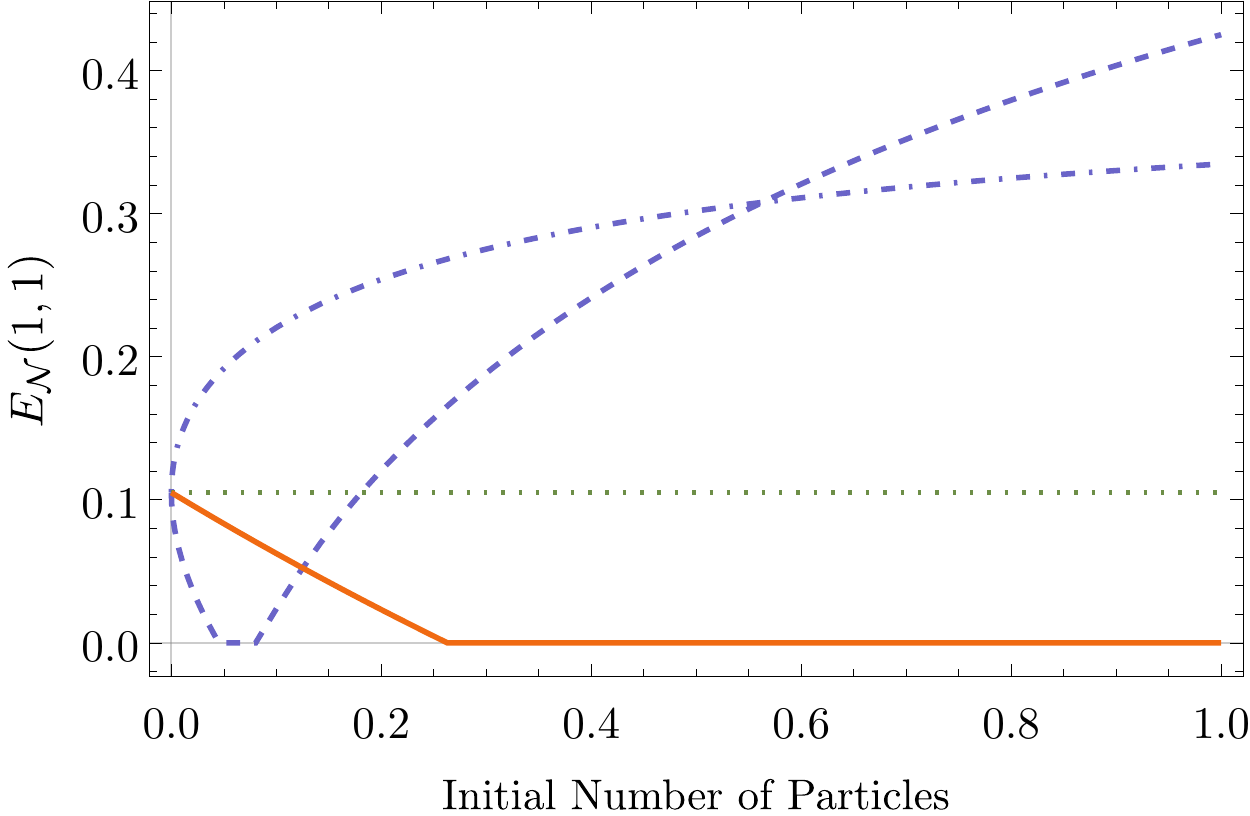}
		\caption{\label{fig:cts_negativity}
			Spatial entanglement of single-mode Gaussian states as a function of their average particle number. Solid, dotted, dashed and dot-dashed lines correspond to thermal~\eqref{eq:init_one_mode}, coherent~\eqref{eq:init_coherent}, and squeezed vacuum~\eqref{eq:init_one_mode} states for $\theta = 0$ and $\theta = \pi/2$, respectively. Plotted is the logarithmic negativity of the reduced state of modes $u_1$ and $\bar{u}_1$, that is the lowest energy modes on the opposite sides of the mirror.}
	\end{figure}
	\subsection{\label{sub:one_mode_squeezed_thermal_states} Single-mode squeezed thermal states} 

	As another input state we consider the first mode of the cavity to be in a squeezed thermal state, while the remaining modes are in their ground states. The vector of first moments and the covariance matrix for this state read as follows:	
	\begin{subequations} \label{eq:init_one_mode}
	\begin{eqnarray}
		\bm{x}^\text{in} &=& 0,\\
		\sigma^\text{in}_{ij} &=& \begin{cases}
			\sigma^\text{in}_{kk} & i = j = 1,\\
			\delta_{ij} \openone & \text{otherwise},
		\end{cases}
	\end{eqnarray}
	\end{subequations}
	where $\delta_{ij}$ is the Kronecker delta and $\sigma^\text{in}_{kk}$ is the covariance matrix of a squeezed thermal state. The latter is characterized by the mean particle number $\bar{n}$, the squeezing coefficient~$s$, the squeezing angle~$\theta$, and is given by
	\begin{widetext}
	\begin{equation} \label{eq:sigma_tilde}
		\sigma^\text{in}_{kk} = \qty(2\bar{n}+1)\mqty(\cosh2s - \cos2\theta \sinh2s & \sin2\theta \sinh2s \\ \sin2\theta \sinh2s & \cosh2s + \cos2\theta \sinh2s).
	\end{equation}
	We note that $\sigma^\text{in}_{kk}$, as given above, is the most general single-mode Gaussian state with vanishing first moments~\cite{WeedbrookEtAl2011}. As previously mentioned, the first moments do not contribute to the amount of spatial entanglement and therefore this state is the most general single-mode Gaussian state for the studies of spatial entanglement harvesting.

	After inserting the above into Eq.~\eqref{eq:sigma_out_om_blocks} and expanding the~$S_{nl}$ blocks, we compute the covariance matrix of the reduced state of the lowest mode on the left and on the right side of the mirror. It has the form $\sigma^\text{out}\vert_{11}$ of Eq.~\eqref{eq:alpha} with
	\begin{equation} \label{eq:sigma_11}
		\sigma_{11} = \bar{\sigma}_{11} = 4 \mathcal{V}_{11}^2 \qty(2\bar{n}+1) \mqty( \omega_1^2 \qty(\cosh2s -\cos2\theta \sinh2s ) & \omega_1 \Omega_1 \sin2\theta \sinh2s \\ \omega_1 \Omega_1 \sin2\theta \sinh2s & \Omega_1^2 \qty(\cosh2s + \cos2\theta \sinh2s) ) + 4 \sum_{l=2}^{2\Lambda} \mathcal{V}_{1l}^2 \mqty(\dmat[0]{\omega_1^2, \Omega_l^2}).
	\end{equation}
	\end{widetext}
	In this case, the expression for~$\gamma_{11}$ is the same as $\sigma_{11}$ with a sign flip at the~$l=2$ term. The reason is that for~$r=R/2$ we have $\mathcal{V}_{12} = -\bar{\mathcal{V}}_{12}$ and $\mathcal{V}_{1l} = \bar{\mathcal{V}}_{1l}$ for $l \ne 2$ [see Eq.~\eqref{eq:nus}].

	\begin{figure}[t]
		\includegraphics[width=0.47\textwidth]{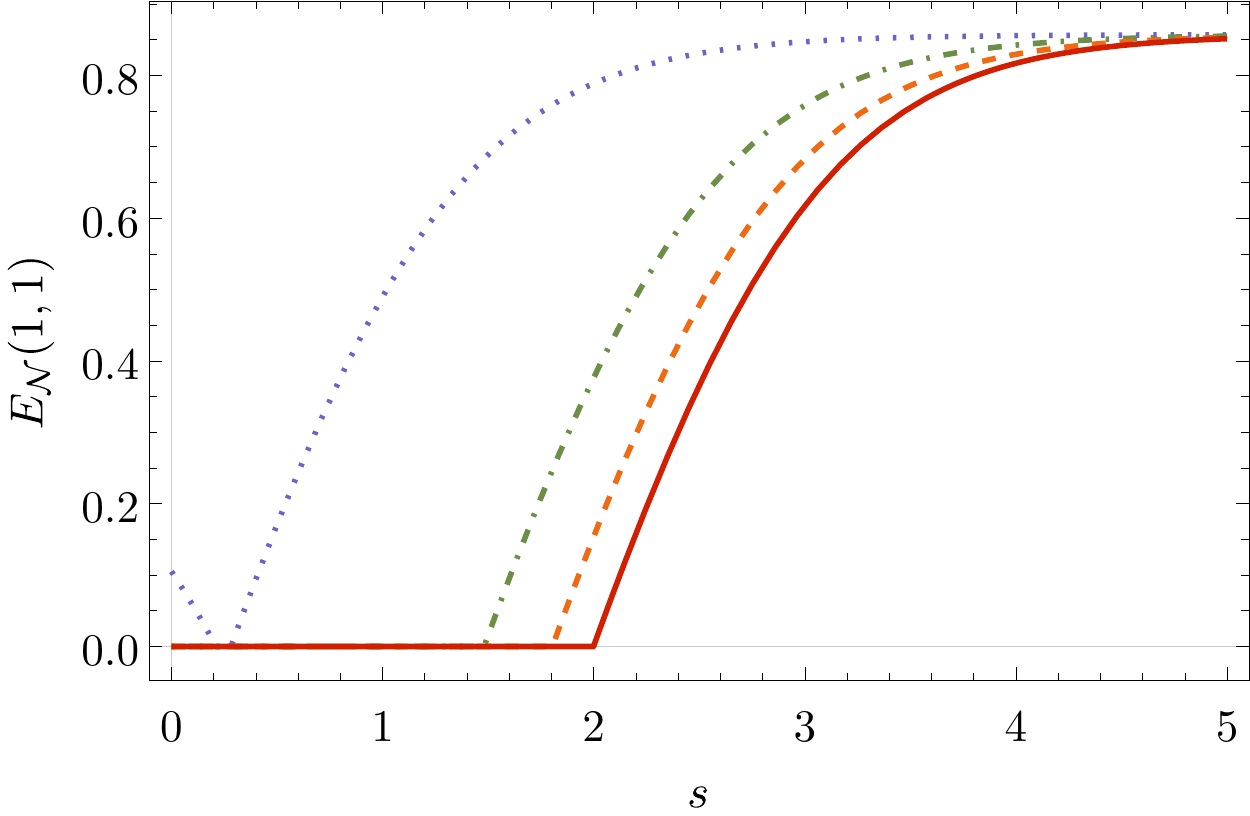}
		\caption{\label{fig:svt}
			Spatial entanglement of the squeezed thermal states~\eqref{eq:init_one_mode} as a function of squeezing coefficient for various temperatures. Dotted, dot-dashed, dashed and solid lines correspond to expectation values of the number of thermal particles $\bar{n} = 0,5,10,15$. The $\bar{n}$ value of 15 approximately corresponds to what could be achieved with a superconducting cavity-on-a-chip setup in dilution refrigerator temperatures~\cite{Note1}.}
	\end{figure}

	Using the reduced state of the modes $u_1$ and $\bar{u}_1$ together with Eq.~\eqref{eq:negativity}, we are able to study the spatial entanglement between these two modes. In Fig.~\ref{fig:cts_negativity}, we have plotted the logarithmic negativity for a number of specific initial states. From this figure, we immediately conclude that the spatial entanglement of nonsqueezed thermal states vanishes rapidly as the temperature of the thermal state increases. As can be observed in the figure, the spatial entanglement is no longer detectable when the expected number of initial thermal particles $\bar{n}$ is greater than approximately $0.2$~\footnote{
		Assuming a cavity length $R = 2\,\mathrm{m}$ and a phase velocity of $c/3$, we can estimate that the temperature corresponding to $\bar{n} = 0.2$ is approximately $1\,\mathrm{mK}$. On the other hand, for a dilution refrigerator temperature of $40\,\mathrm{mK}$ $\bar{n} \approx 15$. The setup parameters we assumed are motivated by currently available superconducting waveguide technology~\cite{BockstiegelEtAl2014}, which is the most promising avenue for experimental realization of our gedanken experiment.}.
	Another conclusion we draw from the results depicted in Fig.~\ref{fig:cts_negativity} is that squeezing generally enhances the amount of extracted spatial entanglement. In particular, as the squeezing parameter of the initial state increases we are able to detect spatial entanglement at higher temperatures (see Fig.~\ref{fig:svt}). We observe that extraction of entanglement from the squeezed thermal state is possible, for any nonzero initial temperature, when the squeezing parameter $s$ exceeds certain threshold. The presence of such threshold has been observed in a general thermodynamical context in~\cite{BruschiNJoP2013,BruschiPRE2015}.

	We now turn to the analysis of the average number of particles produced due to introduction of the extra mirror in the cavity with the initial state given in Eqs.~\eqref{eq:init_one_mode} and~\eqref{eq:sigma_tilde}. Using Eqs.~\eqref{eq:particle_number_formula} and ~\eqref{eq:sigma_11}, we obtain the expectation value of the particle number operator for modes $u_n$ or $\bar{u}_n$: 
	\begin{eqnarray} \label{eq:res_sms_particles}
		\expval{\hat{n}_n} &=& \mathcal{V}_{n1}^2 \qty(2\bar{n}+1) \big[ \qty(\cosh2s -\cos2\theta \sinh2s) \omega_n^2 +\dots\nonumber\\
		&&+ \qty(\cosh2s + \cos2\theta \sinh2s) \Omega_1^2 \big] +\dots\nonumber\\
		&&+ \sum_{l=2}^{2 \Lambda} \mathcal{V}_{nl}^2 \qty(\omega_n^2 + \Omega_l^2) - \frac{1}{2}.
	\end{eqnarray}
	
\begin{figure}[t]
		\includegraphics[width=0.47\textwidth]{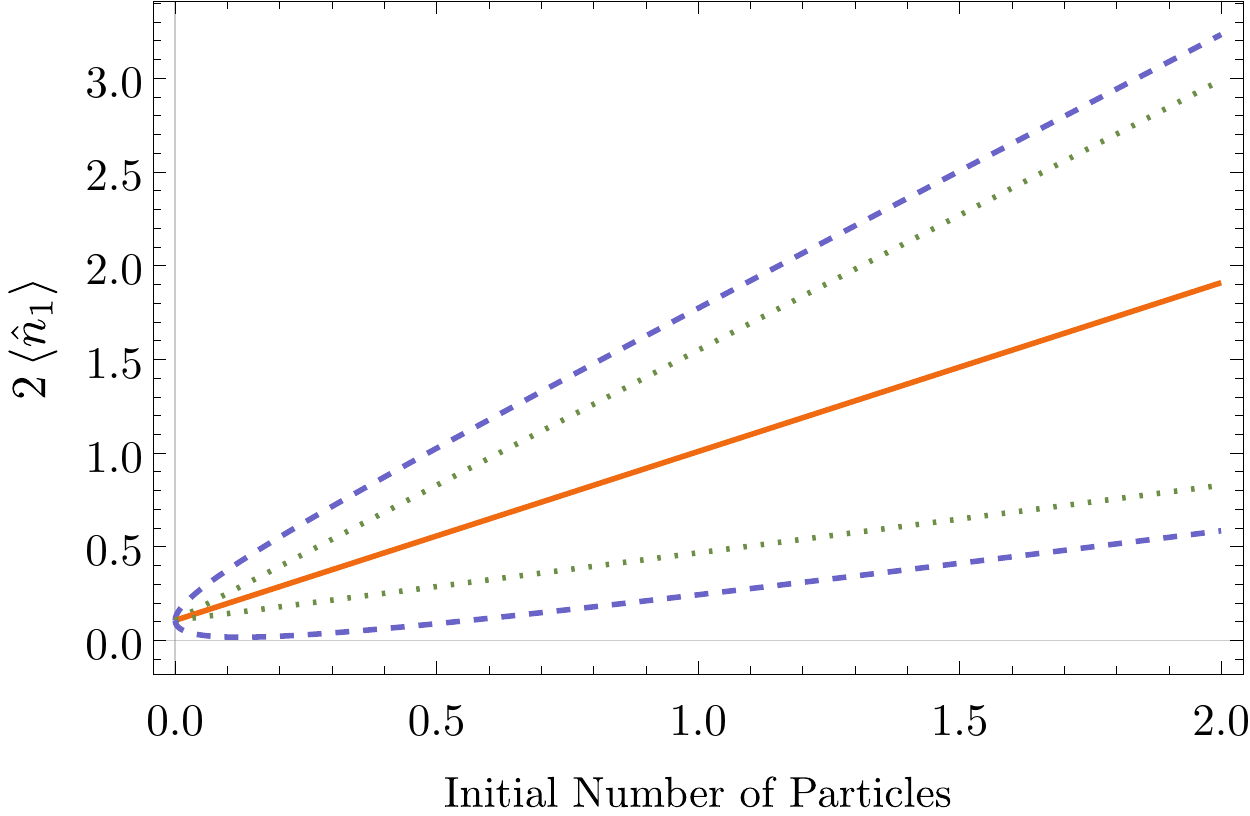}
		\caption{\label{fig:cts_particles}
			Expected number of particles after rapid division of a cavity containing a single-mode Gaussian state as a function of the initial number of particles. Dashed, dotted, and solid lines correspond to the squeezed vacuum~\eqref{eq:init_one_mode}, coherent~\eqref{eq:init_coherent}, and thermal~\eqref{eq:init_one_mode} states, respectively. Only particles produced in the lowest modes energy $u_1$ and $\bar{u}_1$ are shown. The two curves for coherent and squeezed states correspond to the phase and the squeezing angle values of~$0$ and~$\pi/2$.}
	\end{figure}

	In Fig.~\ref{fig:cts_particles}, we have plotted the average number of particles produced in the mode $u_1$, i.e., $\expval{\hat{n}_1}$, for thermal and squeezed states, together with the number of particles~\eqref{eq:res_coherent_particles} that we had previously calculated for initial coherent states. The plot suggests that thermal states yield the same expectation value of the number of particles in the output state as phase-averaged coherent or phase-averaged squeezed states. This is indeed the case, as can be verified by comparing Eqs.~\eqref{eq:res_coherent_particles} and~\eqref{eq:res_sms_particles}. This implies that the values in Fig.~\ref{fig:coh_average_particles} apply also to thermal and to squeezed vacuum states if the squeezing angle can not be controlled.

	\subsection{\label{sub:two_mode_squeezed_vacuum} Two-mode squeezed vacuum} 

	\begin{figure*}[t]
		\includegraphics[width=0.325\textwidth]{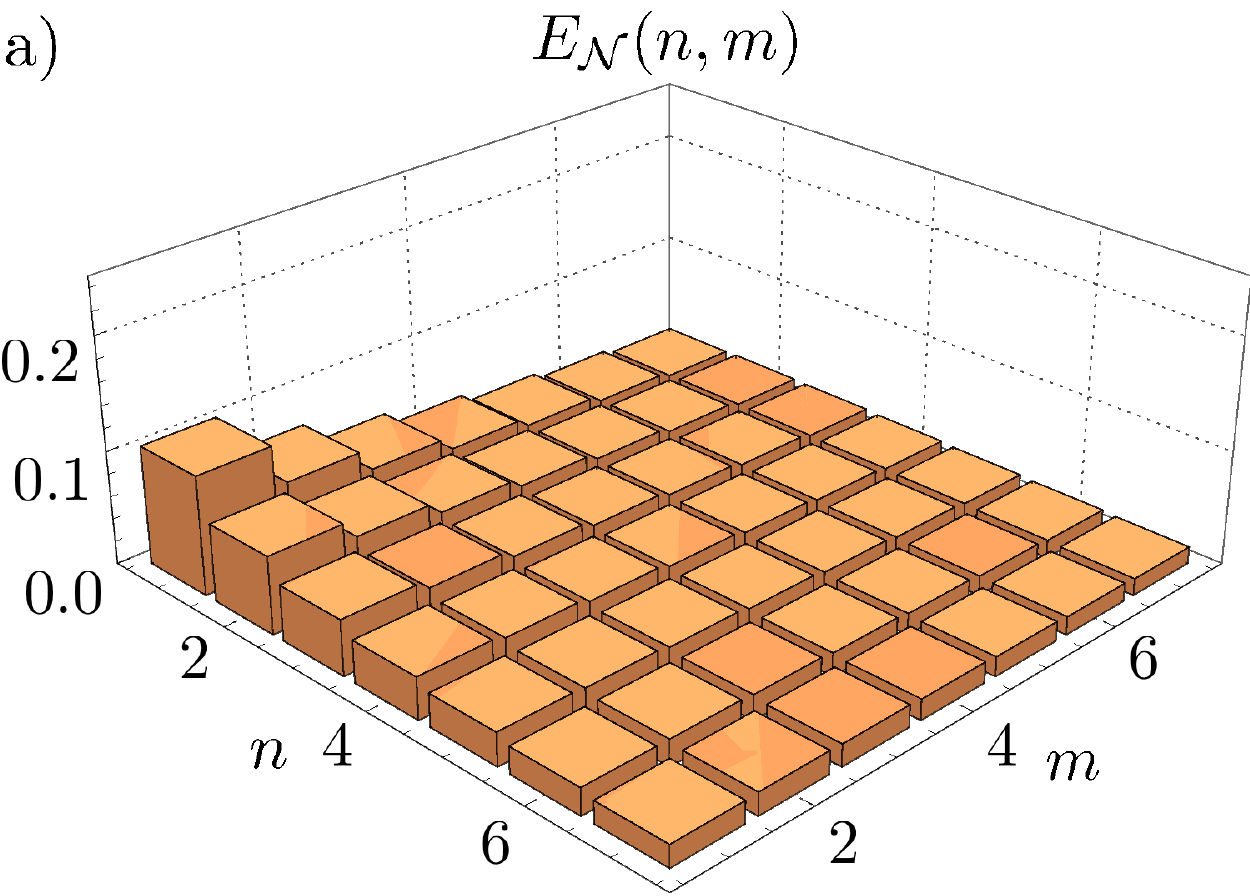}
		\includegraphics[width=0.325\textwidth]{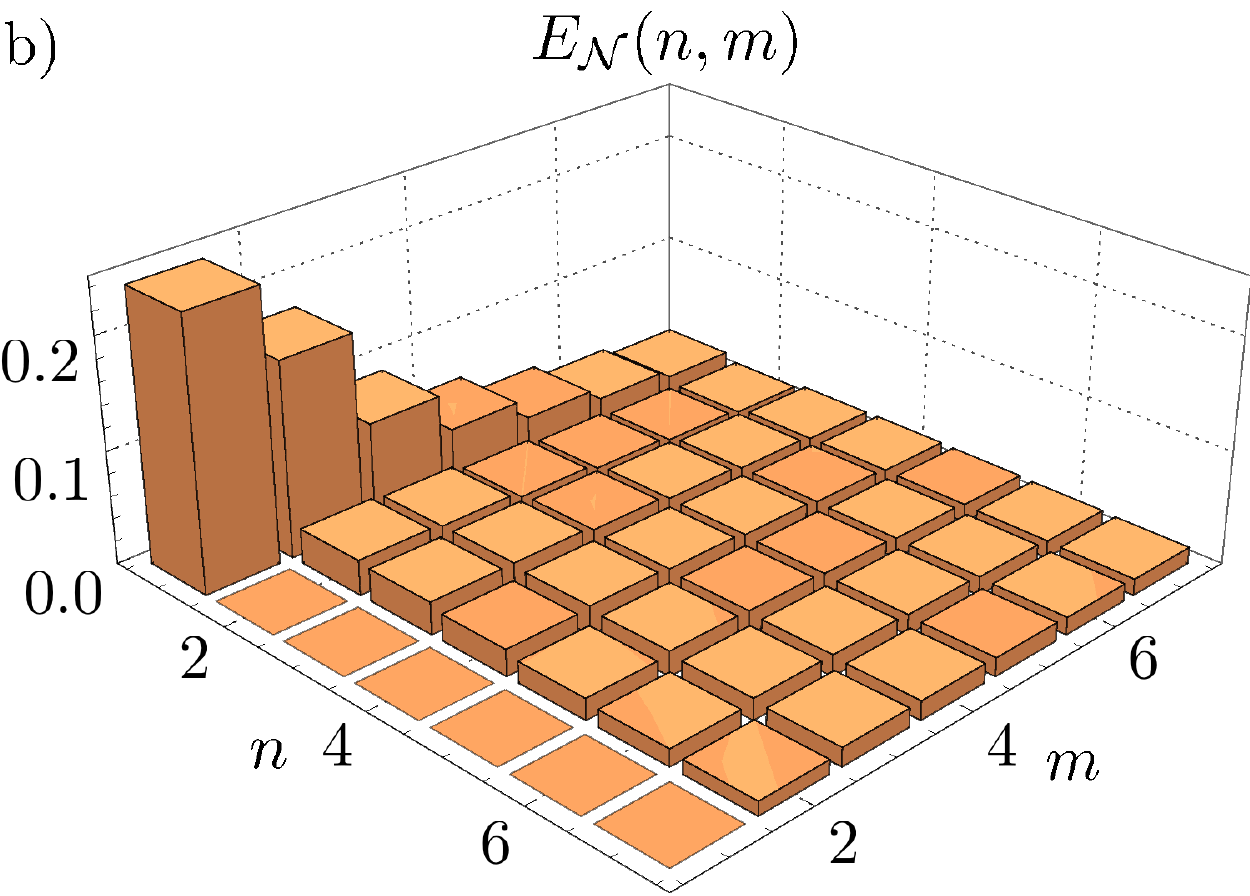}
		\includegraphics[width=0.325\textwidth]{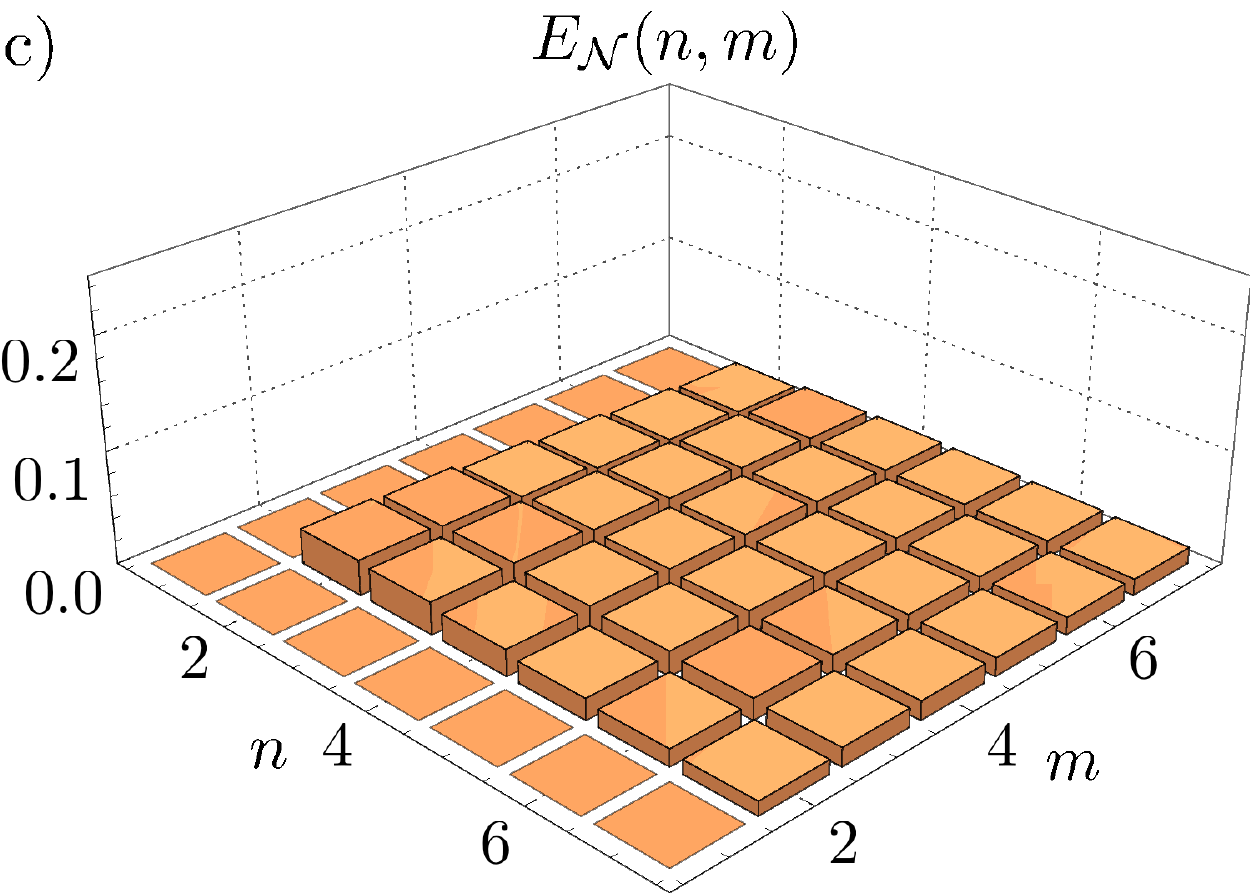}
		\caption{\label{fig:tms}
			Distribution of entanglement between pairs of localized modes $u_n$ and $\bar{u}_m$ [see Eqs.~\eqref{eq:u_n}] for vacuum (a), a two-mode squeezed state (b), and a two-mode squeezed state with the correlations traced out (c). The squeezing parameters are $s = 0.75$ and $\theta = \pi$ [see Eq.~\eqref{eq:init_two_mode_blocks}].}
	\end{figure*}

	In this section, we investigate whether the presence of entanglement between global modes $U_l$ can contribute to the spatial entanglement, i.e., the amount of entanglement in the basis of localized modes $u_n$ and $\bar{u}_m$. To this aim, we compare the spatial entanglement of two states which differ only in the presence of correlations in the initial basis. The first state is a two-mode squeezed vacuum~\cite{WallsMilburn2008} of the modes $U_1$ and $U_2$, which is an entangled state. With $s$ as the squeezing parameter and $\theta$ as the squeezing angle the vector of first moments and the covariance matrix are given by
	\begin{subequations} \label{eq:init_two_mode}
	\begin{eqnarray}
		\bm{x}^\text{in} &=& 0, \\
		\sigma^\text{in}_{ij} &=& \begin{cases}
			\sigma^\text{in}\vert_{12} & i,j \leq 2\\
			\delta_{ij} \openone & \text{otherwise},
		\end{cases}
	\end{eqnarray}
	\end{subequations}
	where the blocks of $\sigma^\text{in}\vert_{12}$, as denoted in Eq.~\eqref{eq:init_two_mode_covm_blocks}, are given by
	\begin{subequations} \label{eq:init_two_mode_blocks}
	\begin{eqnarray}
		\sigma^\text{in}_{11} &=& \sigma^\text{in}_{22} = \cosh(2s) \, \openone, \\
		\sigma^\text{in}_{12} &=& \qty(\sigma^\text{in}_{21})^\mathrm{T} = - \abs{\sinh2s} \mqty( \cos \theta & \sin \theta \\ \sin \theta & -\cos \theta ).
		\end{eqnarray}
	\end{subequations}
	The second state, on the other hand, is a product state built up from the partial traces of the two-mode squeezed state. Therefore, it has the same covariance matrix as the two-mode squeezed state, given by Eqs.~\eqref{eq:init_two_mode} and~\eqref{eq:init_two_mode_blocks}, except that its off-diagonal blocks are zero, i.e., $\sigma^\text{in}_{12} = 0$.

	We first calculate the reduced states of pairs of modes $u_n$, $\bar{u}_m$ for the two initial states introduced above. These reduced states have the form $\sigma^\text{out}\vert_{nm}$ of Eq.~\eqref{eq:alpha} with the $\sigma_{nn}$, $\gamma_{nm}$, and $\bar{\sigma}_{mm}$ blocks given by Eqs.~\eqref{eq:sigma_out_tm_blocks}. Then, to analyze the amount of spatial entanglement, we compute the logarithmic negativity for each mode pair using Eq.~\eqref{eq:negativity} and plot it as a function of $n$ and $m$ in Fig.~\ref{fig:tms}. In this figure, we also include for comparison the known spatial entanglement distribution for the vacuum as the initial state~\cite{BrownEtAl2014}.

	First of all, Fig.~\ref{fig:tms} clearly shows that the entanglement in the basis of $u_n$, $\bar{u}_m$ is very sensitive to the presence of correlations in the initial basis. Second, by comparing the corresponding plots for the vacuum and the two-mode squeezed state, we observe that the presence of correlations in the initial state increases the amount of entanglement between the lowest modes in the final state. Furthermore, the comparison between the two-mode squeezed state and its marginals shows that removing correlations from the initial state, while keeping its partial traces unchanged, causes the entanglement between the lowest modes to vanish.

	Finally, from Fig.~\ref{fig:tms} one can observe that the distribution of spatial entanglement for a two-mode squeezed state can be asymmetric. That is, the mode 2 in the right half of the cavity may be entangled with the mode 1 one the left, while the mode 2 on the left and the mode 1 on the right are separable. However, this does not violate the symmetry between the cavities on left and the right side, as the imbalance can be removed or mirrored to the other side by picking a different value for the two-mode squeezing phase $\theta$.

We would like to conclude this section with a general remark on the dependence of our results on the number of modes taken into account, that is the cutoff number $\Lambda$. We have verified that the state of any single output mode has a well defined $\Lambda \rightarrow \infty$ limit [see for example Eqs.~\eqref{eq:res_coherent_particles}, \eqref{eq:sigma_11}]. In particular, the number of particles in any mode, or entanglement between any mode pair, converges quickly with $\Lambda$. What diverges is the total number of particles summed over all modes of the small cavities. These divergences, however, do not occur if we take into account a finite time of introduction of the mirror. This is because, under such conditions, for modes of sufficiently high frequencies the change in boundary condition is adiabatic, and therefore they remain in their ground states. As a consequence, they do not contribute to the total number of particles. Because of that, the cutoff $\Lambda$ is related to the time scale of the introduction of the additional mirror.

\section{\label{sec:conclusions} Conclusions} 

In this paper, we have revisited the operational approach to study vacuum entanglement given by Brown \emph{et al.}~\cite{BrownEtAl2014}, and applied it to nonvacuum Gaussian states. In this approach a reflecting cavity is divided into two smaller cavities by a rapid introduction of a new mirror. The entanglement between the small cavities is then studied. We have observed that certain nonvacuum states are more effective in revealing spatial entanglement than the vacuum. In particular, we have shown that single-mode and two-mode squeezing typically enhance the amount of spatial entanglement, as can be seen in Figs.~\ref{fig:cts_negativity} and~\ref{fig:tms}. In fact, our results suggest that a degree of squeezing may be indispensable for observation of vacuum entanglement. That is because introduction of squeezing counteracts the increase of temperature (see Fig.~\ref{fig:svt}), which otherwise is detrimental for the spatial entanglement~\cite{Note1}.

We have also demonstrated a link between spatial entanglement of a state and its entanglement between the degrees of freedom that are not spatially localized (see Fig.~\ref{fig:tms}). This provokes interesting and open questions from a resource-theoretic perspective: does adding entanglement between nonlocalized degrees of freedom increase the amount of entanglement between localized observables by the same amount? Can such nonspatial entanglement be extracted using localized observables? Finally, is the reverse process possible, that is the transfer of spatial entanglement to entanglement between, for example, cavity modes of different frequencies?

Last but not least, we have obtained promising results for particle production by sudden introduction of a boundary condition. Figure~\ref{fig:coherent_particles} illustrates that the particle production process is similar to the action of a quantum-optical phase-sensitive amplifier. For in-phase coherent states it suffices to count particles in the two lowest modes of the small cavities to witness an almost twofold increase in the total number of particles. For random-phase coherent states or thermal states, as can be seen from Fig.~\ref{fig:coh_average_particles}, a 14\% increase in the number of particles is still achievable in a similar situation. These results are in strong contrast to what is obtained with the vacuum as the initial state~\cite{Rodriguez-VazquezEtAl2014}, where the number of particles produced is much smaller than any conceivable thermal background. Finally, we would like to point out that similar benefits of nonvacuum initial states have been shown for the shaking-cavity dynamical Casimir effect~\cite{PlunienEtAl2000}.

\begin{acknowledgments}
We thank Christopher Wilson and Iwo Białynicki-Birula for discussions and the National Science Center for financial support from the Sonata BIS Grant No. DEC-2012/07/E/ST2/01402.
\end{acknowledgments}


%

\end{document}